\documentclass[doublecol,linenumbers]{epl2}
\usepackage{multirow} % for 2 columns style with line numbers
\usepackage{cases}

\title{Impact of asymptomatic infection on coupled disease-behavior dynamics in complex networks
}
\shorttitle{Impact of asymptomatic infection on coupled disease-behavior dynamics} %Insert here a short version of the title if it exceeds 70 characters

\author{Hai-Feng Zhang\inst{1,2,4}\and Jia-Rong Xie\inst{5}\and Han-Shuang Chen\inst{3}\and Can Liu\inst{1}\and Michael Small\inst{6,7}
\footnote{michael.small@uwa.edu.au} }

\shortauthor{Hai-Feng Zhang \etal}

\institute{\inst{1} School of Mathematical Science, Anhui University, Hefei 230601, China\\ \inst{2} Center of Information Support \&Assurance Technology, Anhui University, Hefei, 230601, China\\ \inst{3} School of Physics and Material Science, Anhui University, Hefei 230601, China\\
 \inst{4} Department of Communication Engineering, North
University of China, Taiyuan, Shan'xi 030051, China\\ \inst{5} Department of Modern Physics, University of Science and
Technology of China, Hefei 230026, China\\ \inst{6} School of Mathematics and Statistics, The University of Western Australia, Crawley, Western Australia 6009,
Australia\\ \inst{7} Mineral Resources, CSIRO, Kensington, Western Australia, 6151, Australia}

 {Networks and genealogical trees}

\abstract {Studies on how to model the interplay between diseases and behavioral responses (so-called coupled disease-behavior interaction) have attracted increasing attention. Owing to the lack of obvious clinical evidence of diseases, or the incomplete information related to the disease, the risks of infection cannot be perceived and may lead to inappropriate behavioral responses.  Therefore, how to quantitatively analyze the impacts of asymptomatic infection on the interplay between diseases and behavioral responses is of particular importance. In this Letter, under the complex network framework, we study the coupled disease-behavior interaction model by dividing infectious individuals into two states: U-state (without evident clinical symptoms, labelled as U) and I-state (with evident clinical symptoms, labelled as I). A susceptible individual can be infected by U- or I-nodes, however, since the U-nodes cannot be easily observed, susceptible individuals take behavioral responses \emph{only} when they contact I-nodes. The mechanism is considered in the improved Susceptible-Infected-Susceptible (SIS) model and the improved Susceptible-Infected-Recovered (SIR) model, respectively. Then, one of the most concerned problems in spreading dynamics: the epidemic thresholds for the two models are given by two methods. The analytic results \emph{quantitatively} describe the influence of different factors, such as asymptomatic infection, the awareness rate, the network structure, and so forth, on the epidemic thresholds. Moreover, because of the irreversible process of the SIR model, the suppression effect of the improved SIR model is weaker than the improved SIS model.
}

\begin{document}

\maketitle
\section{Introduction} \label{sec:intro}

Many epidemic models have been proposed to enhance our understanding of infectious disease dynamics~\cite{brauer2001mathematical}, however, these mathematical models were often established with static parameters. In reality, outbreak of infectious diseases can trigger the behavioral responses toward diseases, which can further affect the epidemic dynamics. That is to say, the parameters in epidemic models should not be static but dynamic~\cite{funk2015nine}. Therefore, how to establish coupled disease-behavior interaction models to evaluate the interplay between disease dynamics and behavioural responses is becoming a hot field~\cite{funk2010modelling,funk2015nine,wang2015coupled,chen2009modeling,ruan2012epidemic}. There are several key challenges that should be answered in this field~\cite{funk2015nine}: how to incorporate behavioural changes in models of infectious disease dynamics; how to inform measurement of relevant behaviour to parameterise such models; and how to determine the impact of behavioural changes on observed disease dynamics. Along this line, some researchers have already obtained meaningful results. For example, Funk \emph{et al.}~\cite{funk2009spread} have revealed that in a well-mixed population, awareness of epidemics can lead to a lower prevalence of epidemics, but cannot alter the epidemic threshold. Kiss
\emph{et al.} have investigated the impact of information transmission on epidemic outbreaks, and they found that infection
can be eradicated if the dissemination of information is fast enough~\cite{kiss2010impact}.
Perra \emph{et~al.} considered the self-initiated social distancing into classical SIR model, and they found a rich phase space with multiple epidemic peaks and tipping points~\cite{perra2011towards}.

% Meloni \emph{et al.} constructed a meta-population
%model incorporating several scenarios of self-initiated behavioral
%changes into the mobility patterns of individuals, and their results suggest that such behavioral changes
%information do not alter the epidemic threshold.
Previous epidemic models were established in well-mixed populations, however, the transmission of many infectious diseases requires direct or close contact between individuals. As a result, the network-based epidemic models have been extensively investigated~\cite{newman2002spread,pastor2015epidemic}. In particular, studies on how to characterize the interplay between epidemic dynamics and behavioral dynamics within network framework have attracted myriad attention recently~\cite{wang2015coupled,cardillo2013evolutionary,meloni2011modeling}. More importantly, many new and interesting results can be revealed when considered in complex networks. For instance, Refs.~\cite{fu2011imitation,cardillo2013evolutionary,zhang2010hub} have demonstrated that, under voluntary vaccination mechanism, degree heterogeneity
of the network can trigger a broad spectrum of individual vaccinating behavior, where hub nodes are most likely to
choose to be vaccinated --- since they are at greatest risk of infection. Some authors also have shown that local information based behavioral responses can enhance the epidemic threshold and reduce the prevalence of an epidemic~\cite{sahneh2012existence,wu2012impact}, yet, global information based responses cannot alter the epidemic threshold but affect the prevalence of an epidemic~\cite{meloni2011modeling,wu2012impact}. Since the manner of the diffusion of awareness is quite different from the mechanism of epidemic spreading, the coupled disease-behavior interaction models in multiplex networks were also investigated~\cite{granell2013dynamical,wang2014asymmetrically,massaro2014epidemic}. In addition, individuals may change their connections (remove or rewire) when facing the outbreaks of epidemics, so the epidemic dynamics in adaptive networks were considered, and some interesting phenomena, such as assortative degree correlation of evolving network, oscillations, hysteresis and first order transitions can be observed~\cite{gross2006epidemic,zanette2008infection,shaw2010enhanced}.

On one hand, for many infectious diseases, such as H1N1 influenza~\cite{balcan2009seasonal}, severe acute respiratory syndrome (SARS)~\cite{lee2003asymptomatic}, human immunodeficiency virus (HIV)~\cite{parisien1993comparison}, even once individuals have been infected by one kind of disease, they have no evident clinical symptoms, i.e., they are asymptomatic patients; On the other hand, it is difficult for individuals to obtain timely and accurate information related to the diseases. The asymptomatic infection and the incomplete information can affect the behavioral responds towards diseases. Thus, to quantitatively analyze the effects of asymptomatic infection on the epidemic dynamics, in this Letter, we introduce a new compartment---U-state (without evident clinical symptoms, i.e., asymptomatic patients. It is noticed that the U-state is different from the E-state in the standard SEIR model, where E-state is asymptomatic but not infectious, whereas U-state is asymptomatic and infectious) into coupled disease-behavior interaction models. In the model, we assume that susceptible individuals can be infected by the asymptomatic and symptomatic  patients. Nevertheless, since individuals cannot perceive the risks from asymptomatic patients, they alter their behaviors \emph{only} when they contact symptomatic patients. Then consider this mechanism into the improved SIS model and the improved SIR model, respectively. The epidemic thresholds for the two cases are analytically obtained and also verified by numerical simulations. The analytical results can quantitatively describe the impacts of different factors on the epidemic threshold. In addition, we find that the parameters used in models have more significant impact on the SIS model than on the SIR model owing to the reversible process of the SIS model. Our findings may partially answer the key challenges mentioned in the first paragraph proposed in~\cite{funk2015nine}.

%In Sec.~\ref{sec:model}, we describe our two coupled disease-behavior interaction models for the improved SIS and SIR models, respectively. In Sec.~\ref{sec:theory},
%we develop theoretical analyses to understand
%the effects of different behavioral response factors on the spreading dynamics in terms of the
%epidemic threshold and prevalence. Due to
%the intrinsic difference between the SIS and SIR models, we use different theoretical methods to provide the epidemic thresholds. Then the numerical simulations are presented in  Sec.~\ref{sec:result}. Finally,
%we present our conclusions in Sec.~\ref{sec:conclusion}.

\section{Descriptions of the model} \label{sec:model}
\subsection{SAUIS model}

\begin{figure}
\begin{center}
\includegraphics[width=3in]{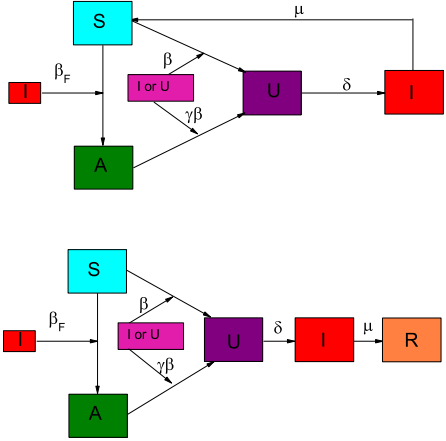}
\caption{(Color online)  Transition diagram for the SAUIS model (upper panel) and the SAUIR model (lower panel). An S-node can be infected by an U- or I-neighbor with rate $\beta$ and becomes an U-node. An S-node can also become aware of infection and goes to A-state with rate $\beta_F$ when contacting an I-neighbor. An A-node can be infected by an U- or I-neighbor with rate $\gamma\beta$ and becomes an U-node. U-nodes become I-nodes with rate $\delta$. For the SAUIS model, the I-nodes recover to S-nodes with rate $\mu$, however, the I-nodes recover to R-nodes with rate $\mu$ and never being infected for the SAUIR model.}
\label{fig1}
\end{center}
\end{figure}

For the classical SIS process in complex networks, where each node in the network can be in one of two states: Susceptible (S) or Infected (I). The infection rate along each SI link is $\beta$, and an infected node can go back to the S-state with a recovery rate $\mu$. In our improved SIS model (named SAUIS model), the I-nodes are divided into two different states: U-state (asymptomatic I-nodes) and I-state (symptomatic I-nodes). All new infected nodes first go to the U-state and then enter the I-state with rate $\delta$. Larger value of $\delta$ means the faster transition from U-state to I-state. Also a new compartment---awareness (A) state is introduced to consider the behavioral responses for the S-nodes. In detail, an S-node can be infected by an U-neighbor or I-neighbor with infection rate $\beta$, and the S-node may go to the A-state when s/he contacts an I-neighbor with awareness rate $\beta_F$. For an A-node, which can also be infected by an U-neighbor or I-neighbor, but with a lower infection rate $\gamma\beta$ with discount rate $0\leq\gamma<1$. I-nodes recover to S-state with recovery rate $\mu$.  The transition diagram is depicted in the upper panel of Fig.~\ref{fig1}.

\subsection{SAUIR model}

Similar to the SAUIS model, an improved SIR model (named SAUIR model) is used to mimic the coupled disease-behavior interaction model. The main difference between the SAUIR model and the SAUIS model is that, for an SAUIR mdoel, I-nodes recover to R-nodes with recovery rate $\mu$ and cannot be infected again. Therefore, the SAUIR model is an irreversible process but the SAUIS model is a reversible process, which yields different theoretical methods to deal with their epidemic thresholds and the results. The transition diagram of the SAUIR model is depicted in the lower panel of Fig.~\ref{fig1}.

\section{Theoretical analysis} \label{sec:theory}
\subsection{Epidemic threshold of the SAUIS model}

Let $S_k(t)$, $A_k(t)$, $U_k(t)$ and $I_k(t)$ be the densities of S-nodes, A-nodes, U-nodes and I-nodes of degree $k$ at
time t, respectively. By using the degree-based mean-field to the SAUIS model, the densities satisfy the following differential equations~\cite{pastor2001epidemic}:
\begin{eqnarray}
 \frac{dS_{k}(t)}{dt}&=&-\beta kS_{k}(\Theta_1+\Theta_2)-\beta_{F}kS_{k}\Theta_2+\mu I_k,\label{11}\\
 \frac{dA_k(t)}{dt}&=&\beta_{F}kS_{k}\Theta_2-\beta\gamma k A_k(\Theta_1+\Theta_2),\label{12}\\
 \frac{dU_{k}(t)}{dt}&=&\beta k( S_{k}+\gamma A_k)(\Theta_1+\Theta_2)-\delta U_k,\label{13}\\
\frac{dI_{k}(t)}{dt}&=&\delta U_k-\mu I_{k},\label{14}
\end{eqnarray}
where $\Theta_1(t)$  and $\Theta_2(t)$ represent the
probabilities that any given link points to an U-node and I-node, respectively. In absence
of any degree correlations, $\Theta_1(t)=\frac{1}{\langle k\rangle}\sum_{k}kP(k)U_{k}(t)$ and $\Theta_2(t)=\frac{1}{\langle k\rangle}\sum_{k}kP(k)I_{k}(t)$ with $P(k)$ being the degree distribution of networks and $\langle k\rangle=\sum_{k}kP(k)$.
The first term in the right side of Eq.~(\ref{11}) accounts for the loss of S-nodes of degree $k$ who are infected by U-neighbors and I-neighbors, and the second term represents that the S-nodes of degree $k$ become A-nodes when contacting I-neighbors, and the third term denotes the recovery of I-nodes of degree $k$. The meanings of the other terms in Eqs.~(\ref{12})-(\ref{14}) can be explained in a similar way.

At the steady state, by imposing Eqs.~(\ref{11}) --(\ref{14}) to be zero, one has
\begin{eqnarray}\label{15}
A_k=\frac{\beta_F\Theta_2}{\gamma\beta(\Theta_1+\Theta_2)}S_k=\frac{\beta_F\delta}{\gamma\beta(\mu+\delta)}S_k
\end{eqnarray}
and
\begin{eqnarray}\label{16}
I_k=\frac{\beta(\mu+\delta)+\delta\beta_F}{\mu\delta}kS_k\Theta_2.
\end{eqnarray}

Since $S_k+A_k+U_k+I_k\equiv1$, by setting $\frac{dI_{k}(t)}{dt}=0$ in Eq.~(\ref{14}), we obtain:
\begin{eqnarray}\label{17}
S_k+A_k+(1+\frac{\mu}{\delta})I_k\equiv1.
\end{eqnarray}

From Eqs. (\ref{15})--(\ref{17}), the following equation can be obtained:

\begin{eqnarray}\label{19}
I_k=\frac{\delta\gamma\beta(\mu+\delta)[\beta(\mu+\delta)+\delta\beta_F]k\Theta_2}{ \mu\delta^2[\gamma\beta(\mu+\delta)+\delta\beta_F]+\gamma\beta(\mu+\delta)^2k\Theta_2[\beta(\mu+\delta)+\delta\beta_F]},
\end{eqnarray}
which induces the following self-consistent Eq.~(\ref{20}):
\begin{widetext}
\begin{eqnarray}\label{20}
\Theta_2=\frac{1}{\langle k\rangle}\sum_{k}kP(k)I_{k}=\frac{1}{\langle k\rangle}\sum_{k}\frac{\delta\gamma\beta(\mu+\delta)[\beta(\mu+\delta)+\delta\beta_F]k^2P(k)\Theta_2}{ \mu\delta^2[\gamma\beta(\mu+\delta)+\delta\beta_F]+\gamma\beta(\mu+\delta)^2k\Theta_2[\beta(\mu+\delta)+\delta\beta_F]}=F(\Theta_2).
\end{eqnarray}
\end{widetext}

%\begin{equation}
%\begin{split}
%\Theta_2&=\frac{1}{\langle k\rangle}\sum_{k}kP(k)I_{k}\\
%&=\frac{1}{\langle k\rangle}\sum_{k}\frac{\delta\gamma\beta(\mu+\delta)[\beta(\mu+\delta)+\delta\beta_F]k^2P(k)\Theta_2}{ \mu\delta^2\gamma\beta(\mu+\delta)+\delta^3\mu\beta_F+\gamma\beta(\mu+\delta)^2k\Theta_2[\beta(\mu+\delta)+\delta\beta_F]}\triangleq F(\Theta_2)
%\end{split}
%\end{equation}
The value $\Theta_2=0$ is always a solution of Eq.~(\ref{20}). In order to has a
non-zero solution, the condition
\begin{eqnarray}\label{21}
\nonumber F'(\Theta_2)\mid_{\Theta_2=0}=\frac{\langle k^2\rangle}{\langle k\rangle}\frac{\gamma\beta(\mu+\delta)[\beta(\mu+\delta)+\delta\beta_F]}{\delta\mu[\gamma\beta(\mu+\delta)+\delta\beta_F]}\geq1
\end{eqnarray}
should be satisfied, where $\langle k^2\rangle=\sum_{k}k^2P(k)$.

By defining $f(\beta)$ as
\begin{eqnarray}\label{22}
f(\beta)=\frac{\gamma\beta(\mu+\delta)[\beta(\mu+\delta)+\delta\beta_F]}{\delta\mu[\gamma\beta(\mu+\delta)+\delta\beta_F]}-\frac{\langle k\rangle}{\langle k^2\rangle},
\end{eqnarray}
the epidemic threshold $\beta_c$ is the point at which $f(\beta)$ just passes through the horizontal when other parameters are fixed (see the insets of Fig.2 and Fig.3).
\subsection{Epidemic threshold of SAUIR model}

For our SAUIR model, on one hand, the epidemic threshold is very difficult or impossible to obtain by solving the mean-field based differential equations; on the other hand, in Ref.~\cite{liu2015interplay}, we have demonstrated that the mean-field method may yield incorrect results when it is used in modified SIR model. Instead, the cavity theory is used to obtain the epidemic threshold for the SAUIR model~\cite{newman2013interacting}.

In our model, during a sufficiently small time interval $dt$, the transition rates of an SU edge becoming an UU edge and an SI edge are $\beta$ and $\delta$, respectively. As a result, the probabilities of an SU edge becoming an UU edge and an SI edge are $T_{11}=\frac{\beta}{\delta+\beta}$ and $T_{12}=\frac{\delta}{\delta+\beta}$, respectively. Similarly, since the transition rates of an SI edge becoming an AI, UI, SR edge are $\beta_F$, $\beta$ and $\mu$, respectively. Therefore, the corresponding probabilities of $SI\rightarrow AI$, $SI\rightarrow UI$ and $SI\rightarrow SR$ are $T_{21}=\frac{\beta_F}{\beta+\beta_F+\mu}$, $T_{22}=\frac{\beta}{\beta+\beta_F+\mu}$ and $T_{23}=\frac{\mu}{\beta+\beta_F+\mu}$. In addition, the probabilities of $AI\rightarrow UI$ and $AI\rightarrow AR$ are $T_{31}=\frac{\gamma\beta}{\gamma\beta+\mu}$ and $T_{32}=\frac{\mu}{\gamma\beta+\mu}$, respectively.

Near the epidemic threshold, the number of infected nodes is very few, statistically, so that each node has one infected neighbor at most. Under such a situation, only the following infection events can happen: 1) an S-node is infected by an U-node or an I-node; or 2) an S-node becomes an A-node and then is infected by the I-node again. However, the probability of an A-node being infected by an U-neighbor is negligible. For an A-node, there must be an I-neighbor to make the S-node become an A-node. According to our statement: each node has one infectious neighbor at most, so there has no U-neighbors again, inducing negligible effect of the probability of $AU\rightarrow UU$. Therefore the infection probability $T$ is given as
\begin{eqnarray}\label{23}
T&=&T_{11}+T_{12}({T_{22}+T_{21}T_{31}})\\
\nonumber&=&\frac{\beta}{\beta+\delta}+\frac{\delta}{\delta+\beta}(\frac{\beta}{\beta+\beta_F+\mu}+\frac{\beta_F}{\beta+\beta_F+\mu}\frac{\gamma\beta}{\gamma\beta+\mu}).
\end{eqnarray}
%The first term denotes an S-node is infected by an U-node (i.e., $T_{11}$). The second term means that the U-node becomes an I-node and then infect then S-node (i.e., $T_{12}T_{22}$),  or first induces an S-node to become an A-node and then infects the A-node again (i.e., $T_{12}T_{21}T_{31}$).

Following the method proposed by Newman \emph{et~al.}~\cite{newman2013interacting}, we first define``externally infected neighbor'' (EIN) for
any node. For a node $i$, if a neighbor $j$ is an EIN then node $j$ is infected by a neighbor other than $i$. Let $\theta$ be the probability that, for any node $i$, whose randomly selected neighbor $j$ is an EIN of $i$. Near the epidemic threshold, the number of infected nodes is very few, indicating that the value of $\theta$ is a very small too. For a node $i$, the excess degree distribution of neighbor $j$ is $q(k)=\frac{(k+1)P(k+1)}{\langle k\rangle}$. Since the probability of each neighbor of $j$ (excluding node $i$) being an EIN is $\theta$ too, and each EIN neighbor of $j$ can infect $j$ itself with probability $T$. Therefore, the probability of $j$ being infected by all EIN is $kT\theta$, then by summing all degree classes with $q(k)$, the probability of node $j$ being the EIN of node $i$ is determined by a self-consistent equation
\begin{eqnarray}\label{24}
\theta=\sum_{k}q(k)kT\theta=\theta TG_1'(1)
\end{eqnarray}
with $G_1(x)=\sum_{k}q(k)x^k$.

From Eq.~(\ref{24}) we have the threshold condition: $TG'_1(1)=1$, which indicates the following equation

\begin{eqnarray}\label{25}
\nonumber&&\frac{\beta}{\beta+\delta}+\frac{\delta}{\delta+\beta}(\frac{\beta}{\beta+\beta_F+\mu}+\frac{\beta_F}{\beta+\beta_F+\mu}\frac{\gamma\beta}{\gamma\beta+\mu})\\&&=\frac{\langle k\rangle}{\langle k^2\rangle-\langle k\rangle}
\end{eqnarray}

should be satisfied for the outbreak of an epidemic.

Similarly, by defining $f(\beta)$ as

\begin{eqnarray}\label{26}
\nonumber &&f(\beta)=\frac{\beta}{\beta+\delta}-\frac{\langle k\rangle}{\langle k^2\rangle-\langle k\rangle}\\&&+\frac{\delta}{\delta+\beta}(\frac{\beta}{\beta+\beta_F+\mu}+\frac{\beta_F}{\beta+\beta_F+\mu}\frac{\gamma\beta}{\gamma\beta+\mu}),
\end{eqnarray}

the epidemic threshold $\beta_c$ is determined by the intersection the curve of $f(\beta)$ and the horizontal  when other parameters are fixed.
\section{Simulation results} \label{sec:result}
In this section, we perform an extensive set of Monte Carlo simulations to validate the theoretical predictions in Section~\ref{sec:model}. Here we implement simulations on configuration networks generated by an uncorrelated configuration model (UCM)~\cite{newman2001random} (We also implemented the models on Erd{\H{o}}s- R{\'e}nyi networks, and found the same results as on the UCM network~\cite{erdHos1959random}). The network contains $N=10000$ nodes and the degree
distribution meets $P(k)\sim k^{-3}$, whose minimal and maximal degrees are $k_{min}=3$ and
$k_{max}=\sqrt{N}$, respectively.

In Ref.~\cite{ferreira2012epidemic}, the susceptibility measure
\begin{eqnarray}\label{measure1}
\chi=N\frac{\langle \rho^2\rangle-\langle \rho\rangle^2}{\langle \rho\rangle}
\end{eqnarray}
is defined by Ferreira \emph{et~al.} to numerically predict the epidemic threshold for the SIS model, where $\rho$ denotes the prevalence of epidemic in one simulation realization. The peak of $\chi$ corresponds to the epidemic threshold.

In Ref.~\cite{shu2015numerical}, Shu \emph{et~al.} proved that the susceptibility measure is not a good measure to determine the epidemic threshold of  SIR model, they then defined the variability measure
\begin{eqnarray}\label{measure2}
\Delta=\frac{\sqrt{\langle \rho^2\rangle-\langle \rho\rangle^2}}{\langle \rho\rangle}
\end{eqnarray}
to numerically determine the epidemic threshold of the SIR model. Their results suggest that the variability measure can predict the epidemic threshold of the SIR model well. In view of this, in this section, we use these two measures to numerically determine the epidemic thresholds of the SIS model and the SIR model, respectively. In our simulations, we have taken at least 1000 independent realizations to predict the epidemic threshold. Without loss of generality, in this study, we set the recovery rate $\mu=1.0$.

\begin{figure*}
\begin{center}
\includegraphics[width=6in]{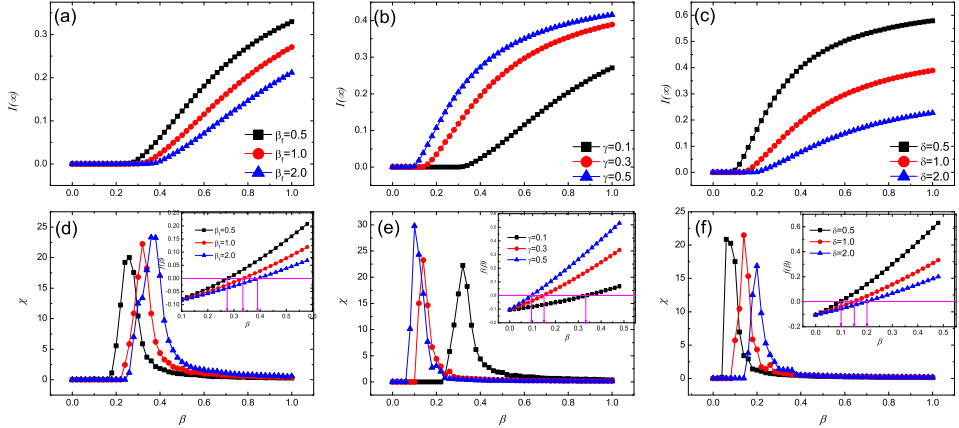}
\caption{(Color online) For SAUIS model, the final epidemic density $I(\infty)$ and the variability measure $\chi$ as functions of $\beta$ in UCM networks. (a) and (d) investigate the impacts of awareness rate $\beta_F$ with $\gamma=0.1$, $\delta=1.0$;  (b) and (e) investigate the impacts of discount factor $\gamma$  with $\beta_F=1.0$, $\delta=1.0$; (c) and (f) investigate the impacts of $\delta$ with $\beta_F=1.0$, $\gamma=0.3$. Insets in bottom subfigures present the theoretical values of $\beta_c$ from Eq.~(\ref{22}). }
\label{fig3}
\end{center}
\end{figure*}

For SAUIS model, the final infected density $I(\infty)$ and the susceptibility measure $\chi$ are plotted as functions of $\beta$ for different cases in Fig.~\ref{fig3}. In general, by comparing the top panels with the bottom panels, one can see that $\chi$ can predict the epidemic threshold $\beta_c$ well. The theoretical values of $\beta_c$ obtained from  Eq.~(\ref{22}) for different cases are given in the insets, which indicate that the theoretical results are in good agreement with the simulation results. Since increasing the awareness rate $\beta_F$ can induce more S-nodes to become A-nodes, as a result, the infection rate is reduced. Therefore, as shown in Figs.~\ref{fig3}(a) and (d),  increasing $\beta_F$ can effectively reduce the final epidemic size and enhance the epidemic threshold. Also, lowering the infection rate of A-nodes (smaller values of $\gamma$) significantly reduces the epidemic size and enhances the epidemic threshold (Figs.~\ref{fig3}(b) and (e)). In particular, the effects of U-nodes on $I(\infty)$ and $\beta_c$ are presented in Figs.~\ref{fig3}(c) and (f). One can observe that the value of $I(\infty)$ is reduced and the value of $\beta_c$ is increased when $\delta$ is increased. With the increase of $\delta$, U-nodes cannot persist for a long time and quickly enter I-state, S-nodes have more chances to become A-nodes because the I-nodes can be easily perceived by S-nodes. Hence, we can understand why increasing the value of $\delta$ can effectively suppress the outbreak of epidemics. We can also conclude that the existence of asymptomatic patients or un-timely information can hinder the behavioral responses of people, which can weaken the suppression effects of behavioral responses on disease controls.

\begin{figure*}
\begin{center}
\includegraphics[width=6in]{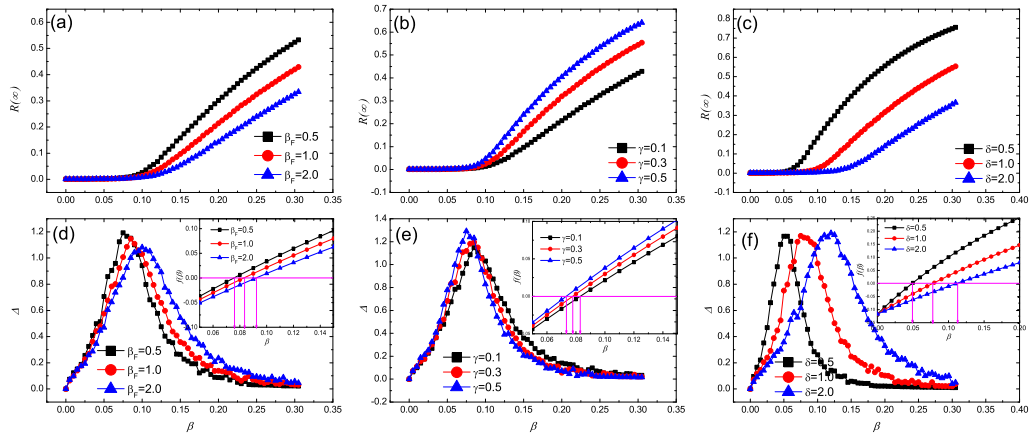}
\caption{(Color online) For the SAUIR model, the final epidemic density $R(\infty)$ and the variability measure $\Delta$ as functions of $\beta$ in UCM networks. (a) and (d) investigate the impacts of awareness rate $\beta_F$ with $\gamma=0.1$, $\delta=1.0$;  (b) and (e) investigate the impact of discount factor $\gamma$  with $\beta_F=1.0$, $\delta=1.0$; (c) and (f) investigate the impact of $\delta$ with $\beta_F=1.0$, $\gamma=0.3$. Insets in bottom subfigures present the theoretical values of $\beta_c$ from Eq.~(\ref{26}).}
\label{fig4}
\end{center}
\end{figure*}

For the SAUIR model, the final infection density $R(\infty)$ and the variability measure $\Delta$ versus transmission rate $\beta$ for different cases are summarized in Fig.~\ref{fig4}. Obviously, the peak of $\Delta$ gives the accurate validation of the epidemic threshold $\beta_c$, and the theoretical values from Eq.~(\ref{26}) (the insets) are in accordance with the numerical results. Moreover, just like the results in Fig.~\ref{fig3}, increasing the values of $\beta_F$ and $\delta$, or reducing the value of $\gamma$ can reduce the final epidemic size and enhance the epidemic threshold. Nevertheless, by comparing Fig.~\ref{fig3} with Fig.~\ref{fig4}, we find that the suppression of behavioral responses on epidemics for the SAUIR model is worse than the case of the SAUIS model, especially for the impact of $\beta_F$ (Figs.~\ref{fig4}(a) and (d)) and $\gamma$ (Fig.~\ref{fig4}(b) and(e)).  As for the SAUIR model, the epidemic process ends quickly owing to its irreversible nature of the model, which causes individuals have no sufficient time to take behavioral responses.

\section{Conclusions} \label{sec:conclusion}

In this Letter we have studied the coupled disease-behavior interaction model in complex networks by dividing infectious individuals into asymptomatic (U-state) and symptomatic individuals (I-state). Then the epidemic thresholds for the improved SIS model and SIR model were obtained by using different theoretical methods. The analytic results for the epidemic thresholds can \emph{exactly} show how great impacts of U-state, network structure, awareness rate, and so forth, have on the epidemic dynamics. In addition, because of the irreversible process of the SAUIR model, the suppression effect of behavioral responses on disease control is not as good as the case of the SAUIS model. Our findings provide a typical example to emphasize the importance of incorporating human behavioral response into epidemic models, and also partially offer a theoretical tool to quantify the impacts of behavioral responses.
\begin{acknowledgments}
This work is funded by the National Natural Science Foundation of China
(Grant Nos. 61473001, 11205002, 11331009). MS is funded by an Australian Research Council Future Fellowship (FT110100896).
\end{acknowledgments}
%\bibliographystyle{eplbib}
%\bibliography{Spreading}

\end{document}